\documentclass{sig-alternate}
\pdfoutput=1

\def \imagedir {.}
\input{glyphtounicode}
\usepackage[utf8]{inputenc}
\usepackage[T1]{fontenc}

\usepackage[english]{babel}
\usepackage{graphicx}
\usepackage{caption}
\usepackage{subcaption}
\usepackage{balance}  				
\usepackage{listings}					
\usepackage{fixltx2e}					

\newcommand{\ie}{i.\,e.}
\newcommand{\eg}{e.\,g.}
\newcommand{\wrt}{w.\,r.\,t.~}
\newcommand\circa{{\raise.17ex\hbox{$\scriptstyle\mathtt{\sim}$}}}

\def\sharedaffiliation{%
\end{tabular}
\begin{tabular}{c}}
\begin{document}


\title{Dynamic Physiological Partitioning\\on a Shared-nothing Database Cluster}
\numberofauthors{2}
\author{
	\alignauthor
	Daniel Schall\\
  \email{schall@cs.uni-kl.de}
	\alignauthor
	Theo H\"arder\\
  \email{haerder@cs.uni-kl.de}
  \sharedaffiliation
   \affaddr{Databases and Information Systems Group}  \\
   \affaddr{University of Kaiserslautern, Germany}
}

\maketitle
\begin{abstract}
Traditional DBMS servers are usually over-provisioned for most of their daily workloads and, because they do not show good-enough energy proportionality, waste a lot of energy while underutilized.
A cluster of small (wimpy) servers, where its size can be dynamically adjusted to the current workload, offers better energy characteristics for these workloads.
Yet, data migration, necessary to balance utilization among the nodes, is a non-trivial and time-consuming task that may consume the energy saved.
For this reason, a sophisticated and easy to adjust partitioning scheme fostering dynamic reorganization is needed.
In this paper, 
we adapt a technique originally created for SMP systems, called \texttt{physiological partitioning}, to distribute data among nodes, that allows to easily repartition data without interrupting transactions.
We dynamically partition DB tables based on the nodes' utilization and given energy constraints and compare our approach with \texttt{physical partitioning} and \texttt{logical partitioning} methods.
To quantify possible energy saving and its conceivable drawback on query runtimes, we evaluate our implementation on an experimental cluster and compare the results w.r.t. performance and energy consumption.
Depending on the workload, we can substantially save energy without sacrificing too much performance.
\end{abstract}

\section{Introduction}
Saving energy is a concern in all areas of IT.
Studies have shown that single servers have potential for energy-optimizations, but in general, the best performing configuration is also the most energy efficient one \cite{DBLP:conf/sigmod/TsirogiannisHS10}.
This observation stems from the fact that the power spectrum between idle and full utilization of a single server is narrow and $\circa 50$\% of its power is already consumed at idle utilization \cite{DBLP:journals/computer/BarrosoH07}.

Today's server hardware is not energy proportional, hence, at low utilization, the hardware---mainly main memory and storage drives---consume a significant amount of power.
Therefore, better energy efficiency cannot be achieved with current, centralized hardware solutions.
This observation also holds for traditional DBMSs, composed of a single server with huge main memory and lots of storage drives attached.
In contrast to centralized, brawny servers, a scale-out cluster of lightweight (wimpy) servers 
is able to shutdown single nodes independently.
At an abstract level, this enables the cluster to dynamically add storage and processing power based on the cluster's utilization.
Similar to cloud-based solutions, we hypothesize that a cluster of nodes may adjust the number of active (power consuming) nodes to the current demand and, thus, approximate energy proportionality.

Based on these observations, we developed WattDB, a research prototype of a distributed DBMS cluster, running on lightweight, Amdahl-balanced nodes using commodity hardware.
The cluster is intended to dynamically shrink and expand its size, dependent on the workload.
Although the cluster may not be as powerful as a monolithic server, for typical workloads, we expect our system to consume significantly less energy.

Reconfiguring a cluster to dynamically match the workload requires data to be 
redistributed among the active nodes to balance the utilization.
Yet, copying data is time-consuming and adds overhead to the already loaded cluster.
Reducing both, time and overhead, is crucial for an elastic DBMS.

In this paper, we adapt \emph{physiological partitioning}, proposed by Pinar T{\"o}z{\"u}n et al. \cite{DBLP:journals/vldb/TozunPJA13} to a distributed database, benchmark it against other partitioning approaches, and show further optimizations.
After giving an overview of recent research addressing  partitioning, elasticity, and energy efficiency of DBMSs in Sect.~\ref{section:RelatedWork}, we discuss important operational aspects of 
\emph{WattDB}.
In Sect.~\ref{section:Partitioning}, we introduce our adaptation of \emph{physiological partitioning} and compare it to \emph{physical} and \emph{logical partitioning}.
Sect.~\ref{section:Experiments} contains the results of several empirical experiments for dynamically reorganizing DB clusters using one of the three techniques.
In Sect.~\ref{section:Conclusion}, we emphasize some important observations concerning the use of partitioning in a dynamic DB cluster. 

\section{Related Work}
\label{section:RelatedWork}
Reducing energy consumption of servers as well as dynamic reconfiguration and efficient DB partitioning are all subject to various research papers.
In the following, we give a short overview of related works in the three fields that serve as building blocks of our research.

\subsection{Database Partitioning}
Partitioning a table is an old concept and widely used.
Splitting tables into multiple partitions has mainly two advantages:

First, by dividing the table into smaller logical groups---either by key ranges, hash, or based on time intervals---the amount of data necessary to access for a particular query can be reduced.
All major DBMSs use techniques like \emph{partition pruning} or \emph{partition-wise joins} to reduce the amount of data to be read for a query \cite{Oracle11gPartitioning}.
DB partitioning also enables parallelization, and thus, better utilization of the hardware and higher performance by parallelizing data accesses.
This is especially true for a distributed DBMS, where partitions can be allocated to various nodes, enabling processing on more CPU cores and---in contrast to single-node databases---also bringing in more MMUs\footnote{MMU = Memory Management Unit, providing additional bandwidth to resolve the bottleneck between CPU and main memory.}, main memory, and storage disks to support query processing.

Second, partitions can be used as units of logical control over the data contained, \ie, the node owning a partition is responsible for its integrity and concurrency control.
By dividing a large table into smaller partitions, the resulting control overhead can be shared among nodes.
Hence, instead of a single node having to manage an entire table with long request queues waiting for locks, 
to perform a variety of integrity checks, and to serialize processing due to log writes, all tasks can be split up into partitions and maintained by a group of nodes.

Partitions provide a logical encapsulation over a group of records, \ie, records do not span multiple (horizontal) partitions.
Hence, moving a partition from one node to another does not affect other parts of the table.
For this reason, DB partitioning is an ideal candidate to provide the building block for dynamic reorganization in our DBMS. 

T{\"o}z{\"u}n  et al. proposed a \textit{physiological partitioning} scheme in the context of a multi-threaded DBMS (\cite{DBLP:journals/vldb/TozunPJA13,pandis2010data}).
They identified two existing techniques, physical and logical partitioning.
Whereas physical partitioning is equivalent to the data distribution of a shared-nothing DBMS, logical partitioning corresponds to a shared-everything approach.
In their work, they introduced multi-rooted B*-trees, each identifying a partition of the table.
By allowing only a single thread at a time to access such a tree, they eliminated contention and locking overhead.
While they focused on a single node with multiple CPU cores to assign partitions to, we are implementing a similar partitioning technique, but focus on \textit{physically distributed nodes to store partitions}.

\subsection{Dynamic Clustering}
Traditional clustered DBMSs do not dynamically adjust their size (in terms of the number of active nodes) to their workload.
Hence, scale-out to additional nodes is typically supported, whereas the opposite functionality, shrinking the cluster and centralizing the processing---the so-called scale-in---, is not.
Recently, with the emergence of clouds,  a change of thinking occurred and dynamic solutions became a research topic.

In his PhD thesis \cite{Das:2011:SET:2521552}, Sudipto Das implemented an elastic data storage (called \emph{Elastras}) able to dynamically grow and shrink on a cloud.
As common in generic clouds, his work is based on decoupled storage, hence, all I/O involves network communication.
He introduced \emph{Key Groups}, an application-defined set of records which are frequently accessed together.
These groups can be seen as dynamic partitions that are frequently formed and dissolved.
By distributing the partitions among nodes in the cluster, both performance and cost can be controlled.

Our work does not target traditional clouds, but rather a cluster of nodes whose hardware properties are well known and with dedicated storage disks connected directly to each node.
Yet, a similar implementation could be running on cloud platforms.

\subsection{Energy Optimizations}

Lang et al. \cite{Lang:2012:TED:2350229.2350280} have shown that a cluster suffers from ``friction losses'' due to coordination and data shipping overhead and is therefore not as powerful as a comparable heavyweight server.
On the other hand, for moderate workloads, \ie, the majority of real-world DB applications, a scale-out cluster can exploit its ability to reduce or increase its size sufficiently fast and, in turn, gain far better energy efficiency.

We already explored the capabilities and limitations of a clustered storage architecture \cite{Schall:2013A} that dynamically adjusts the number of nodes to varying workloads consisting of simple \emph{read-only page requests} where a large file had to be accessed via an index\footnote{Starting our WattDB development and testing with rather simple workloads facilitated the understanding of the internal system behavior, the debugging process, as well as the identification of performance bottlenecks.}.
We concluded that it is possible to approximate energy proportionality in the storage layer with a cluster of wimpy nodes. However, attaching or detaching a storage server is rather expensive, because (parts of) datasets may have to be migrated. Therefore, such events (in appropriate workloads) should happen on a scale of minutes or hours, but not seconds.

In \cite{Schall:2013B}, we have focused on the query processing layer---again for varying workloads consisting of two types of \emph{read-only SQL queries}---and drawn similar conclusions.
In this paper, we revealed that attaching or detaching a (pure) processing node is rather inexpensive, because repartitioning and movement of data is not needed.  Hence, such an event can happen in the range of a few seconds---without disturbing the current workload too much.

We substantially extended the kind of DBMS processing supported by WattDB to \emph{complex OLAP / OLTP workloads consisting of read-write transactions} in \cite{SH-DASFAA2014}.
For this purpose, we refined and combined both approaches to get one step closer to a fully-featured DBMS, able to process OLTP and OLAP workloads simultaneously.
In this work, we were able to trade performance for energy savings and vice versa.
Yet, we identified that  cluster adaptation  and  data distribution to fit the query workload is time-consuming and needs to be optimized.

\section{Energy-Aware Database Cluster}
As discovered before, a single-server DBMS is far from being energy proportional and cannot process realistic workloads in an energy-efficient way.
Our previous research indicates that a cluster of lightweight (wimpy) servers, where nodes can be dynamically switched on or off, seems more promising.

In single-server-based solutions, the best performing configuration is usually also the most energy-efficient one (see \cite{DBLP:conf/sigmod/TsirogiannisHS10}).
In a clustered environment, on the other hand, increasing the number of nodes might improve overall query performance, but---due to the increased power consumption---without lowering energy consumption per query.

Further, a dynamic cluster of nodes heavily relies on repartitioning to re-allocate data  and, in turn, equally balance utilization among all nodes. 
Otherwise, hotspots or bottlenecks on a single node would slow down the entire cluster and, thus, lead to bad performance and energy figures.


\subsection{Power Consumption}
Our cluster consists of n (currently 10) identical nodes, interconnected by a Gigabit Ethernet.
Each node is equipped with an Intel Atom D510 CPU, 2 GB of DRAM, and three storage devices: one HDD and two SSDs.
The configuration is considered Amdahl-balanced, 
 \ie, balanced w.r.t. I/O and network throughput on one hand and processing power on the other.
By choosing commodity hardware with limited data bandwidth, GB-Ethernet wiring is sufficient for interconnecting nodes.
All nodes can communicate directly.

Each wimpy node consumes \circa22 -- 26 Watts when active (based on utilization) and \circa2.5 Watts in standby.
The interconnecting network switch consumes 20 Watts and is included in all measurements.
In its minimal configuration---with only one node and the switch running and all other nodes in standby---the cluster consumes \circa65 Watts.
This configuration does not include any disk drives, hence, a more realistic minimal configuration requires \circa70 -- 75 Watts.
In this state, a single node is serving the entire DBMS functionality (storage, processing, and cluster coordination).
With all nodes running at full utilization, the cluster will consume \circa260 to 280 Watts, depending on the number of disk drives installed.
This is another reason for choosing commodity hardware which uses much less energy compared to server-grade components.

\subsection{DBMS Software}
By the time, research gained interest in energy efficiency of DB servers, no state-of-the-art DBMS was able to run on a dynamically changing cluster.
To test our hypotheses (see Section 1), we developed a research prototype called \emph{WattDB} that supports traditional query processing with ACID properties, but is also able to dynamically adjust to the workload by scaling out or in, respectively.
The smallest configuration of WattDB is a single server called \emph{master node}, hosting all DBMS functions and always acting as the cluster coordinator and endpoint to DB clients\footnote{Note, this imposes a single point of failure.}. To relieve this node, DB objects (tables, partitions) and query evaluation can be offloaded to 
arbitrary cluster nodes.

\subsection{Dynamic Query Processing}
In order to run queries on a cluster of nodes, distributed query plans are generated on the master node.
Almost every query operator can be placed on remote nodes, excluding data access operators which need local access to the DB records.
Running those operators on remote nodes would increase access times without additional gains.

Running query operators on a single node does not involve network communication among query operators, because all records are  transferred via main memory.
Distributing operators implies shipping of records among nodes and, hence, introduces network latencies.
Additionally, the bandwidth of the Gigabit Ethernet, which we are using for our experiments, is relatively small, compared to memory bandwidth.

To mitigate the negative effects of distribution, WattDB is using \emph{vectorized volcano-style query operators} (\cite{Graefe:1994:VEP:627290.627558,conf/cidr/BonczZN05}), hence, operators ship a set of records on each call.
This reduces the number of calls between operators and, thus, network latencies.
To further decrease network latencies, buffering operators are used to prefetch records from remote nodes.
\emph{Buffering operators} act as  proxies between two (regular) operators; they asynchronously prefetch records, thus, hiding the delay of fetching the next set of records.

In WattDB, the query optimizer tries to put pipelining operators\footnote{Pipelining operators can process one record at a time and emit the result, \eg, projection operators.} on the same node to minimize latencies. Offloading pipeline operators to a remote node has little effect on workload balancing and, thus, does not pay off.
In contrast, blocking operators\footnote{Blocking operators need to fetch all records from the underlying operators first, before they can emit the first result record, \eg, sorting operators.} may be placed on remote nodes to equally distribute query processing.
Blocking operators generally consume more resources (CPU, main memory) and are therefore good candidates for offloading and hence, balancing utilization in the cluster.

\begin{figure*}[t]
	\centering
	\begin{minipage}[b]{.49\textwidth}
		\includegraphics[width=0.9\textwidth]{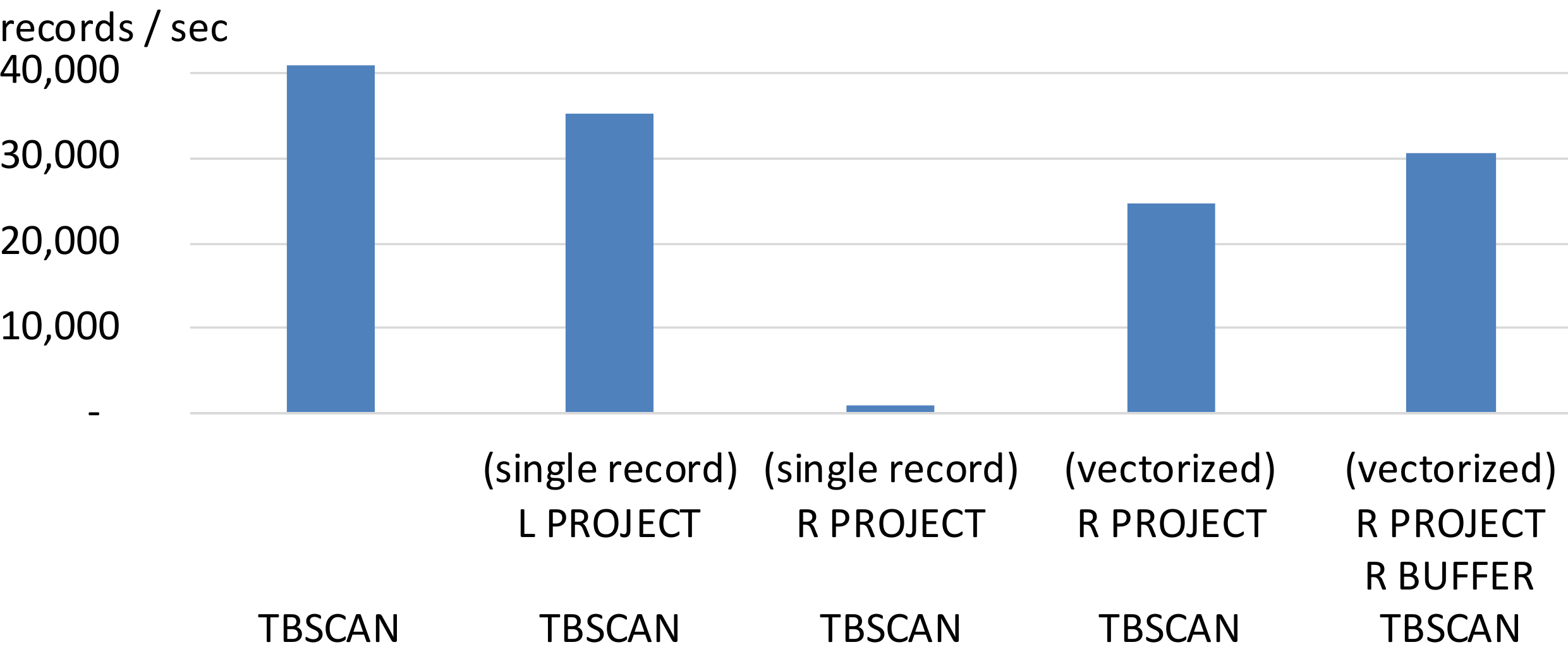}%
		\caption{Micro-benchmark testing record throughput}%
		\label{figure:offloading}%
		\vspace{-0.3cm}
	\end{minipage}\hfill
	\begin{minipage}[b]{.49\textwidth}
		\includegraphics[width=0.9\textwidth]{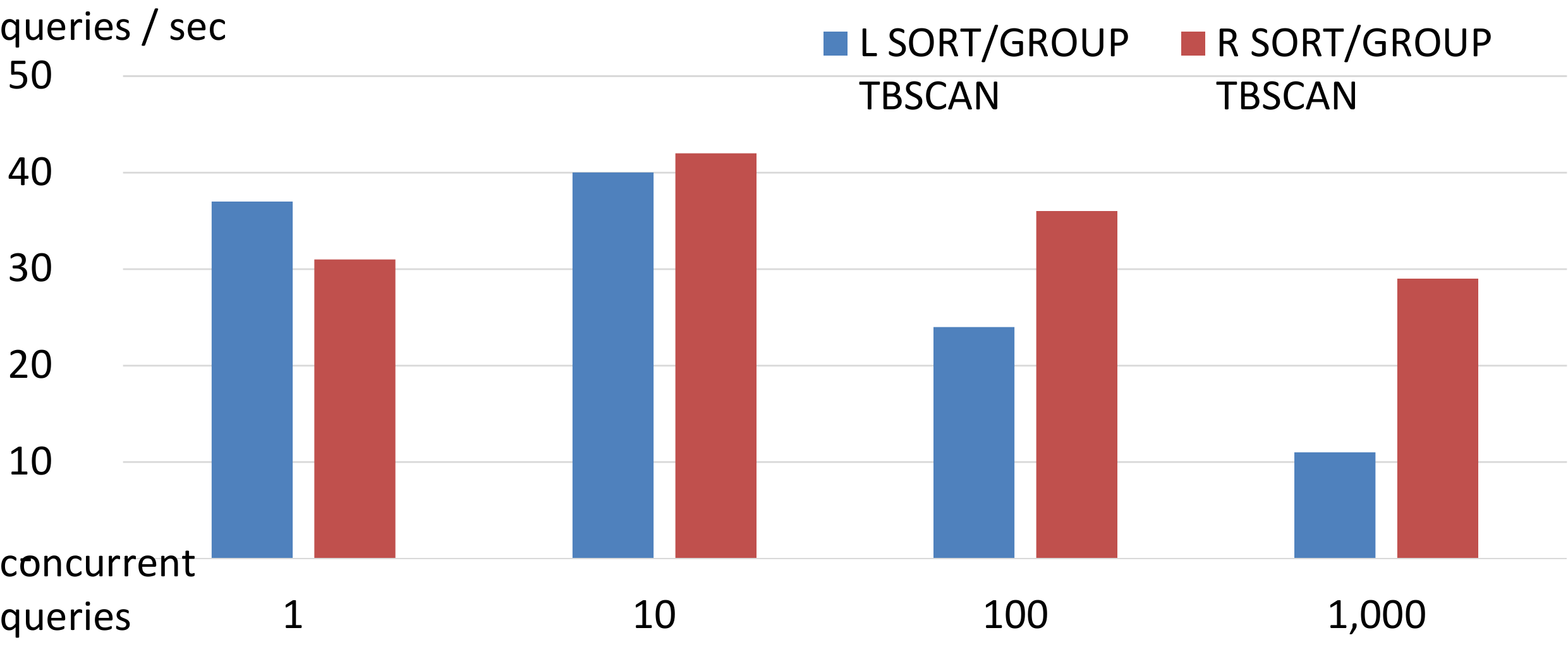}%
		\caption{Offloading queries, throughput}%
		\label{figure:overloading}%
		\vspace{-0.3cm}
	\end{minipage}
\end{figure*}

In Fig. \ref{figure:offloading}, we show results of a micro-benchmark, demonstrating the performance impact of distributing operators among nodes.
The first (leftmost) run is a query containing a  table scan locally running on a single node.
The maximum throughput is slightly more than 40,000 records per second.
In the next run, we added a local projection operator on top of the table scan, running on the same node.
Although classic volcano-style operators  ship only one record at a time,
throughput is still high (approx. 34,000 records per second).
To identify the influence of distribution, we ran the same operator combination on remote nodes.
In this setting, throughput drops to less than 1,000 records per second, because each call to \texttt{next()} involves network delays.
Next, we run the same operators on remote nodes with vectorized operators, hence, each call to \texttt{next()} returns a set of records at once.
As a result, the operators need less calls to fetch all records and throughput increases to 24,000 records per second.
In the last experiment, we included a buffering operator, which runs on the remote node and prefetches results from the underlying table scanner.
While the projection operator is still processing a set of records, the buffer operator can asynchronously prefetch new records to further minimize network delays.
In this constellation, throughput further increases to \circa30,000 records per second.

Hence, with vectorized, volcano-style operators, network delays can be successfully minimized to allow almost arbitrary operator placement in the cluster.
Looking at these results, it is quite evident that distributing queries, instead of running all operators locally, is always a performance burden.
Although we could reduce the negative impact, local query processing is still faster, compared with all other measurements.
This is true for isolated queries, running on an underutilized node.

Yet, in a typical DBMS, multiple queries compete for resources like buffer space and CPU cycles.
In these cases, offloading parts of the query plan to another node and, thus, reducing the node's utilization may even improve performance.
To verify that offloading query operators to other nodes can increase overall query throughput, we have designed another micro-benchmark.
For this experiment, we have run multiple queries concurrently, each query consists of a table scan with a subsequent sorting phase.
In Fig. \ref{figure:overloading}, the throughput is shown for varying numbers of concurrent queries.
The left (blue) bars plot the throughput for a single-node query plan, where both operators run on the same node.
With increasing parallelism, throughput drops, because the node is overloaded and the queries compete for CPU and buffer.
On the right side (red bars), the sorting operator is off\-loaded to another node.
Because of additional network communication, query throughput is initially lower than in the all-local case.
With more concurrent queries, the additional buffer space and CPU power pay off, and query throughput becomes substantially  higher, compared to the single-node case.

With these (simple) experiments, we validate that distributing queries among nodes may increase overall performance, despite the additional network delays.
Still, careful considerations have to be made regarding the available network bandwidth and the nodes' utilization to estimate whether or not offloading will pay off.
Also, offloading queries at low utilization levels is inferior to centralized processing.

\subsection{Dynamic Reorganization}
The master node is coordinating the whole cluster.
It is globally optimizing the query plans, whereas individual nodes can locally optimize their part of the plan. Furthermore, it takes nodes on- and offline and decides when and how the tables are (re)partitioned.

Every node is monitoring its utilization: CPU, memory consumption, network I/O, and disk utilization (storage and IOPS).
Additionally, 
performance-critical  data is collected for each DB partition, \ie, CPU cycles, buffer page requests and network I/O.  With these figures, we can correlate the observed utilization of the cluster components to the  (logical) DB entities.
Hence, both types of data are necessary to identify sources of over-utilization.
We use the performance figures of the components to identify their over- or underutilization. In addition, the monitoring of the activities of the DB entities are needed to determine the origin of the cluster's imbalance.
For this reason,
the nodes send their monitoring data every few seconds to the master node.

The master checks the incoming performance data to predefined thresholds---with both upper and lower bounds.
If an overloaded component is detected, it will decide where to distribute data and whether to power on additional nodes and resume their cluster participation.
Similar, underutilized nodes trigger a scale-in protocol, \ie, the master will distribute the data (processing) to fewer nodes and shutdown the nodes currently not needed.
Decisions, what data to migrate and where, are done based on the current utilization of the nodes, the expected query workload, and the estimated cost, it will take to migrate data between nodes.

In WattDB, we have implemented different policies regarding the scale-out behavior.
First, each node in the cluster stores data on local disks to minimize network communication.
If a node goes out of storage space, DB partitions are split up on nodes with free space. 

Second, WattDB tries to keep the I/O rate for each storage disk in a certain range.
Underutilized disks are eligible for additional data---either newly generated by INSERT operations or migrated from overloaded disks.
Utilization among storage disks is first locally balanced on each node, before an allocation of data from/to other nodes is considered.

Third, each node's CPU utilization should not exceed the upper bound of the specified threshold (80\%).
As soon as this bound is violated for a node, WattDB first tries to offload query processing to underutilized nodes\footnote{This works well for operators like \textbf{SORT}, \textbf{GROUP}, and \textbf{AGGREGATE}.}.
In case, the overload situation cannot be resolved by redistributing the query load, the current data partitions and their node assignments are reconsidered.
If a partition causing the CPU's overload is identified, it is split according the partitioning scheme applied, whereupon affected records are moved to another node.
The exact details of data redistribution vary, based on the partitioning scheme chosen.
In case of underutilized nodes, 
a scale-in protocol is initiated, which quiesces the involved nodes from query processing and shifts their data partitions to nodes currently having sufficient processing capacity. 

Similar rules exist for network and memory utilization, but in the experiments performed in this paper, they were never triggered.
WattDB makes decisions based on the current workload, the course of utilization in the recent past, and the expected future workloads \cite{KRAMER12}.
Additionally, workload shifts can be user-defined to inform the cluster of an expected change in utilization.

\subsection{Multiversion Concurrency Control}
Dynamic re-allocation of data should have as little impact as possible on the services provided by the cluster, while ACID quality for the DB operations is maintained. For this reason, all record movement operations have to be protected from multi-user anomalies, whereupon  "real" transaction throughput should be kept as high as possible. 
Therefore, concurrency control should  impede data access and operation as little as possible---even when data is \textit{on the move}. So-called system transactions are provided to guarantee serializability of record movement \cite{Graefe:2011}. 

\emph{Multiversion Concurrency Control} (MVCC) allows multiple versions of DB objects to exist, modifying a record creates a new version of it without deleting the old one immediately.
Hence, readers can still access old versions, even if new transactions changed the data.
This property is especially useful for dynamic partitioning techniques, where records are frequently moved, \ie, deleted and re-created on another partition.

To select the best-suited concurrency control mechanism, we have conducted a series of micro-benchmarks comparing classical Multi-Granularity Locking with RX lock modes (MGL-RX)  with Multiversion Concurrency Control (MVCC) \cite{Bernstein:1983:MCC:319996.319998}.\footnote{For details of the implementation of MVCC in WattDB, see \cite{SH-DASFAA2014}.}
\begin{figure}
	\centering
	\includegraphics[width=0.47\textwidth]{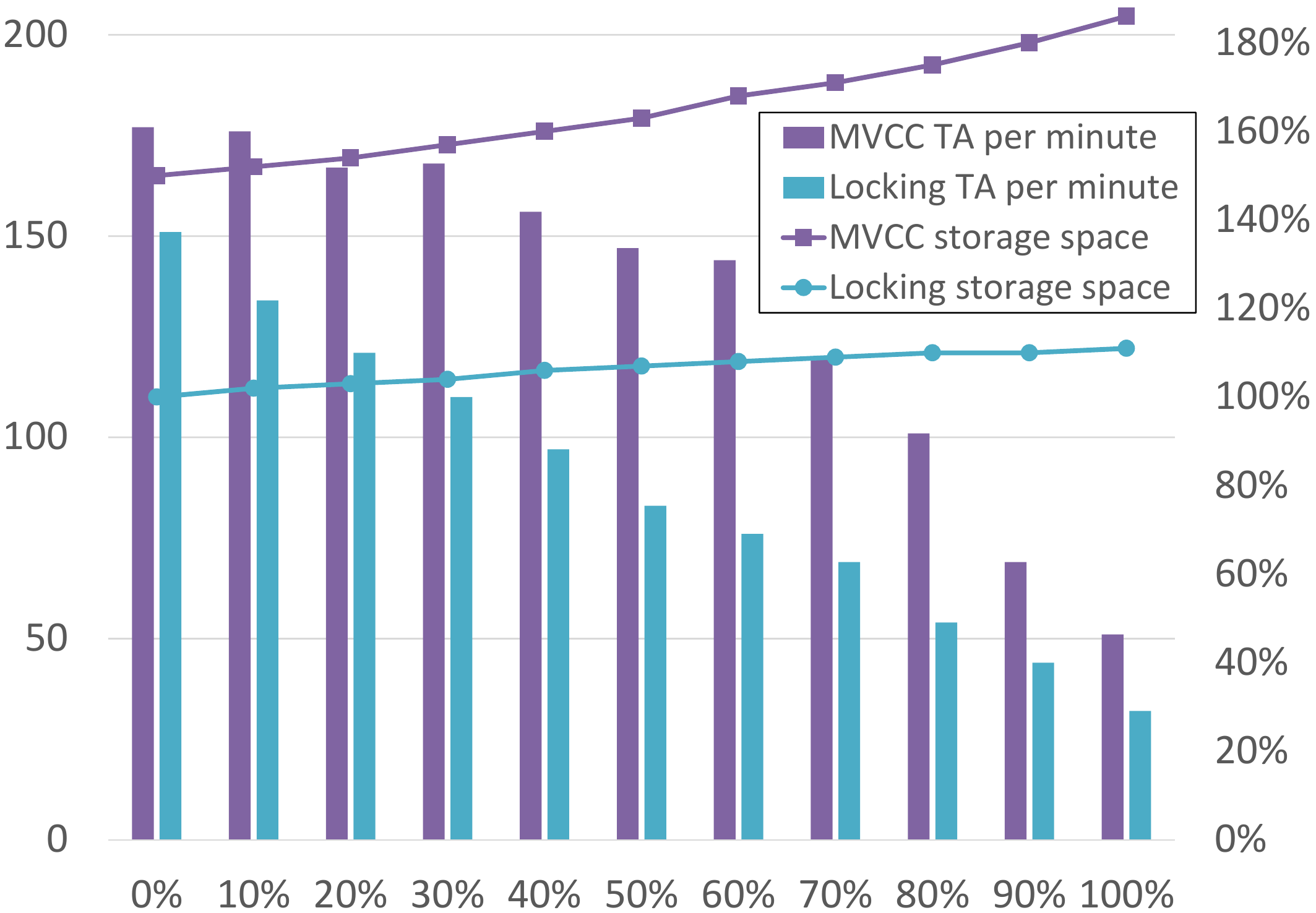}
	\caption{MVCC vs. MGL-RX: performance and storage space consumption of workloads with different amount of updates while moving records}
	\label{figure:mvcc:move}
\end{figure}
We have compared the performance of MGL-RX with MVCC, while moving 50\% of the records to another partition.
To get an impression of MVCC's potential for our application, we have measured the performance and storage requirements for different ratios of read-only and write-intensive transactions (see Fig. \ref{figure:mvcc:move}).
The X-axis shows the percentage of update transactions, whereas the remaining percentage is that of read-only transactions.
The graph bars depict the query throughput using MVCC and MGL-RX, respectively.
The lines show the storage requirements for both mechanisms.

The experiment shows that MVCC can increase transaction throughput between 15\% (for read-only workloads) and almost 90\% (for pure writer workloads), while the affected partition is moved.
Storage requirements for MVCC are obviously higher, as multiple versions of records have to be kept.
Traditional locking also requires additional storage space to hold a list of pending changes, which have to be applied to the data after their move is finished.

\section{Dynamic Partitioning}
\label{section:Partitioning}
\begin{figure}[t]%
	\centering
		\includegraphics[width=\columnwidth,page=2]{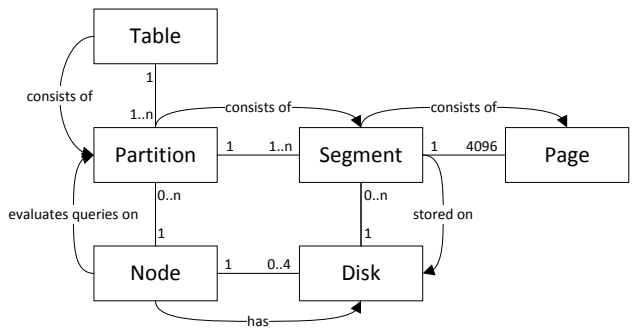}%
		\vspace{-0.3cm}
		\caption{Database schema}%
		\label{figure:schema}%
		\vspace{-0.3cm}
\end{figure}
In order to scale a cluster of nodes dynamically to the workload, it is necessary to repartition the database online.
Nodes still having data on disk must not shut down to prevent data loss or inaccessibility.
A flexible, fast, and ACID-compliant partitioning scheme ensures that data can still be accessed even while repartitioning takes place.

First, we need to clarify the terms we are using in this paper:

\textbf{Tables:}
A DB \emph{table} is a purely logical construct in WattDB.
Its metadata (column definitions, partitioning scheme) is maintained on the master node.
Each table is composed of k \emph{horizontal partitions}\footnote{\url{http://dev.mysql.com/tech-resources/articles/partitioning.html} outlines horizontal partitioning.}, each belonging to a specific node, responsible for query evaluation, data integrity (logging), and access synchronization (locking).
The partitioning scheme used is application-depen\-dent, as some applications may benefit from distinct key ranges, while others may prefer scattered data.

\textbf{Partitions:}
Each partition contains 1 to m \emph{segments}, which are physical units of storage.
Each segment is located on a specific disk on a node in the cluster.
Segments stored on the same node as the partition do not require network access to fetch data, but accessing only local disks may impose an I/O bottleneck for the partition.
Therefore, it is also possible to remotely address segments, stored on other nodes.
Assignment policies of segments to partitions depend on the partitioning scheme used, \ie, whether to allow segments to be stored on remote nodes (shared disk) or not (shared nothing).
Based on their assignment to nodes and disks, access costs to segments vary.
Partitions are by default index-organized \cite{Srinivasan:2000:OIT:645926.672004} \wrt the primary key with support for additional, secondary indexes.

\textbf{Indexes:}
In WattDB, \emph{indexes} are realized using B*-trees and span only one partition at a time.
Hence, indexes are stored on the same partition as the data and do not contain cross-references to other partitions.

\textbf{Segments:}
A segment (32 MB) consists of 4096 \emph{blocks} or \emph{pages}, which are consecutively stored on disk.
Segments are the unit of distribution in the storage subsystem. Hence, all pages in a segment will be copied/moved among nodes in one batch.
The data granularity inside the buffer is a page, which is also the unit of data transfer between nodes.

Fig. \ref{figure:schema} clarifies these terms 
and their relationships.
Using our experimental prototype, we are now ready to evaluate and compare several partitioning approaches.

\begin{figure}[t]%
  \begin{subfigure}[t]{0.9\columnwidth}
		\includegraphics[width=\textwidth,page=1]{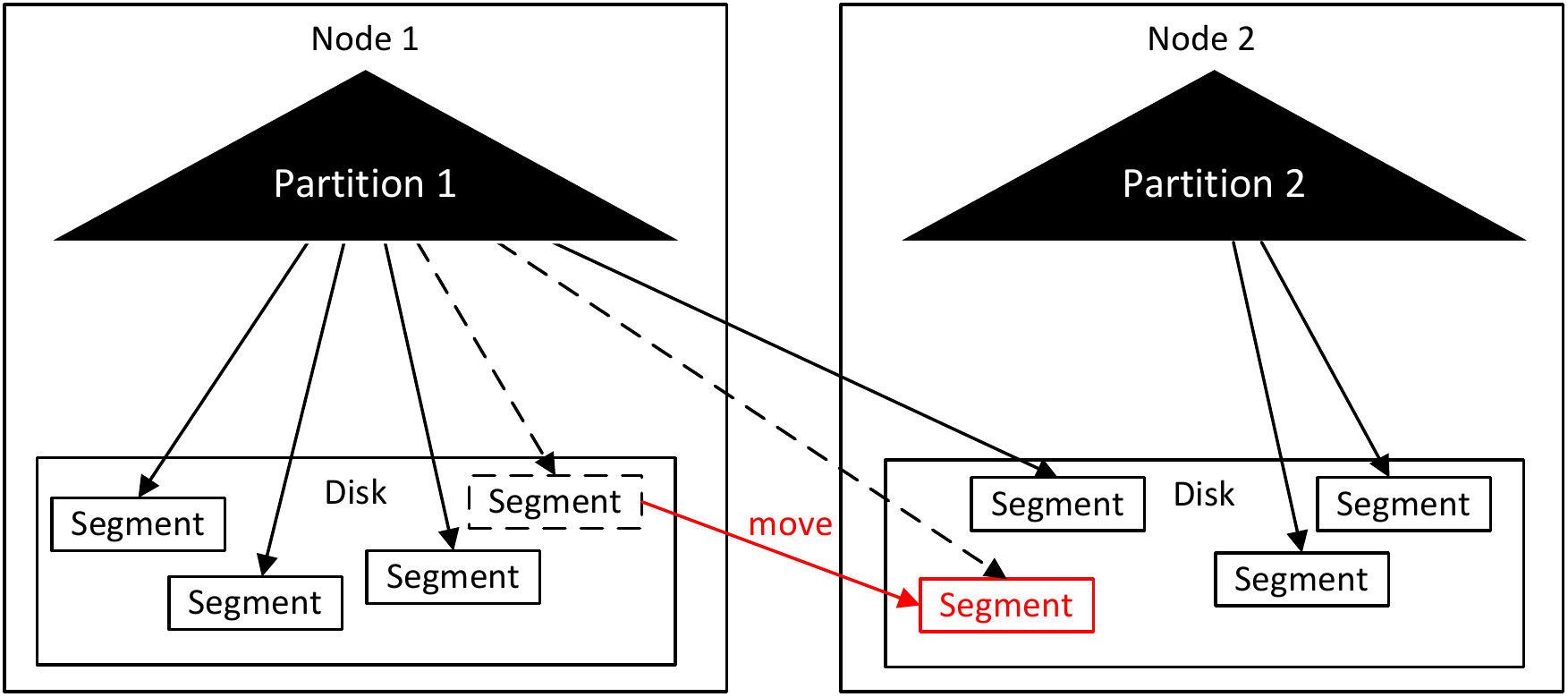}%
		\caption{Physical partitioning}%
		\label{figure:partitioning:physical}%
	\end{subfigure}\\
  \begin{subfigure}[t]{0.9\columnwidth}
	\vspace{0.2cm}
		\includegraphics[width=\textwidth,page=2]{\imagedir/Partitioning}%
		\caption{Logical partitioning}%
		\label{figure:partitioning:logical}%
	\end{subfigure}\\%
  \begin{subfigure}[t]{0.9\columnwidth}
	\vspace{0.2cm}
		\includegraphics[width=\textwidth,page=3]{\imagedir/Partitioning}%
		\caption{Physiological partitioning}%
		\label{figure:partitioning:physiological}%
	\end{subfigure}%
	\caption{Illustration of the partitioning schemes considered}%
	\label{figure:partitioning schemes}%
	\vspace{-0.5cm}
\end{figure}

\subsection{Physical Partitioning}
\emph{Physical partitioning} operates at the data access layer and does not change logical access paths.
The logical DB layer is oblivious of the segment distribution in the storage layer.
To repartition, whole segments are moved among nodes, without altering the data stored inside.
Fig. \ref{figure:partitioning:physical} depicts physical partitioning.
A shared-everything architecture on the storage layer is needed to support physical partitioning among nodes, hence, every server needs to be able to access every segment, local and remote.

At the logical layer, segments are exclusively assigned to nodes, independent of their disk placement to ensure integrity and eliminate the need for coordination.
Hence, cluster nodes will not share data stored in segments; 
for logical data access, it behaves like a shared-nothing architecture.

Logical access paths remain unchanged and keep pointing to the same page addresses while repartitioning, even when the physical placement of segments (and thus, pages) changes.
A mapping between logical and physical page addresses is needed.
This approach is easy to achieve and does not require extensive housekeeping.
With physical partitioning, data can be easily spread out to multiple disks and hence, increase IOPS and access bandwidth.
Since logical entities are  not affected, movement is transparent to higher DB layers and does not need any changes in access paths.
The partition structure is not altered, hence, the primary-key ranges of the table's partitions remain unchanged and  need no updates to reflect the new partitioning.

While this is an advantage of physical partitioning, it is also a drawback:
Physical partitioning affects only the storage layer by distributing segments to disks/nodes.
The query execution layer does not benefit from additional nodes hosting the data, because the logical control remains on the original node.
Therefore, the query optimizer is unaware of the changes in the physical layer.
Also, placing segments on remote nodes induces network access latencies, multitudes higher than local-disk access latencies.
The intermediate network may also induce a bandwidth bottleneck.

Transactions are not needed to physical repartitioning, because logical records are not accessed;
a lightweight latching/synchronization mechanism, locking segments on the move for a short time, is sufficient.

From an energy-concerned perspective, spreading data out to additional disks on remote nodes increases power consumption without adding much query performance.

\subsection{Logical Partitioning}
In contrast, \emph{logical partitioning} moves records from one partition to another and, hence, affects the logical DB layer.
To balance the workload among partitions, records within key ranges are moved between nodes.
This requires the use of transactions to guarantee ACID properties:
Records are removed from one partition and inserted into another;
transactions need to ensure that concurrent transactions read either copy, but not both.
Fig. \ref{figure:partitioning:logical} shows an example of logical partitioning, where each partition holds records from a distinct primary-key range.
All segments of a partition are stored locally, \ie, on the same node to minimize network latency, hence, logical partitioning can be implemented on a shared-nothing architecture.\footnote{It is also possible to implement logical partitioning on shared disk, but for the sake of simplicity, in this paper, logical partitioning implies shared nothing.}

While rebalancing, dedicated transactions delete records in one partition and insert them into another.
Hence, data movement alters the key ranges of the partitions.
The query optimizer can take the new partitioning for future query optimization into account.
Spreading data over multiple partitions, stored on separate nodes, increases IOPS and bandwidth on the physical layer due to the increased number of disks---as with physical partitioning.
Further, by logically dividing the data into key ranges, stored separately, the query optimizer can prune unneeded partitions.
Additionally, using logical partitioning, ownership of the records changes, and all nodes holding segments can access partitions in parallel and thus, speed up query processing.

Yet, to remove records with a specific key range from a partition, a large part of the data must be read and updated, possibly scattered among physical pages.
Hence, logical partitioning is more IO-heavy than physical partitioning.
Since transactions are needed, queries running in parallel may get delayed due to locking conflicts.

\subsection{Physiological Partitioning}
Because both methods sketched so far have drawbacks, we have extended \emph{physiological partitioning} \cite{DBLP:journals/vldb/TozunPJA13} 
to partition data among nodes, not CPU cores.
Similar to their original approach, we encapsulate key ranges in partitions and assign them exclusively.
While the authors of physiological partitioning assign partitions to CPU cores to eliminate contention, we assign partitions to nodes for the same reason.

Partitions still consist of segments, but each segment keeps a primary-key index for all records within it.
Hence, partitions only contain an index on top, keeping information about key ranges in the attached segments.
This top index is very small compared to an index containing all records from all segments.
Moving a segment from one partition to another does not invalidate the primary-key index of the segment.
To reflect the changes in the partitioned DB, only an update to both of the top indexes (of the new and old partition) is required.

Fig. \ref{figure:partitioning:physiological} sketches the design of physiological partitioning.
Two partitions on different nodes are shown, both consisting of several sub-partitions, contained in segments.
Primary-key ranges for each of the mini-partitions are outlined.

To repartition the table, it is sufficient to move whole segments, containing mini-partitions, to another node.
The receiving node can immediately resume query processing, while old transactions may still finish reading from the old segment on the sending node.
New queries will already access the  segment on the new node.

Like physical partitioning, physiological partitioning copies data almost at raw disk speed.
Additionally, the logical layer is aware of the new data distribution and can participate in query processing as with logical partitioning, \eg, the query optimizer can perform \emph{segment pruning}, allowing a query to quickly identify unnecessary segments, having no interesting data.
Also, buffering, synchronization, and integrity control for the segment are now transferable to another node---not possible with physical partitioning.
Using physiological partitioning, we can still apply MVCC for concurrency control.

\textbf{Repartitioning details:} Rebalancing the DB cluster, exemplified by the movement of a single segment, works as follows:
First, the partition is marked for repartitioning on the master node and the partition tree on the source node is updated with a pointer to the new location of the partition.
Next, on the source node, a read lock is acquired on the source partition, waiting for pre-existing queries to finish updating the partition.
Updating transactions need to commit before the lock is granted.
By ensuring that all changes to the partition are committed, no UNDO information needs to be shipped to another node.

After the lock is granted, the partition is copied to the target node and inserted into the node's partition tree.
At this point, the new partition is unlocked and records in it can be accessed by readers and writers again.
The master node is informed of the successful movement operation, and the global partition table is updated accordingly.
New transactions will now be redirected to the new node directly.

The partition information on the source node still points to the target node, redirecting all queries trying to access the old partition to the new one.
Finally, after all old transactions no longer want to access the old partition, the master informs the old node to unlock the partition.
At that time, the pointer to the new node is removed from the source node, and the old partition can safely be removed.

\textbf{Logging:} For durability reasons, write-ahead logs must be maintained at all times.
When repartitioning, although record ownership changes, log files remain on the original node and are not transferred to the node hosting the partition.
In case of DB failures, the log file is needed to reconstruct partitions and to perform appropriate UNDO and REDO operations.
Since moving a partition involves read-locking the entire partition, this operation acts as a checkpoint.
All transactions before the movement will have their actions recorded in the old log file.
While moving the partition, old copies of the records still remain until the movement is finished.
Hence, additional logging is not required.

After successfully moving a partition to another node, the partition will be in a consistent state and flushed to disk.
Hence, the old copies and the old log file are no longer required.
Now, updates to the new partition can be logged on the new node.

\textbf{Housekeeping on the master:} Query optimization is done on the master node.
To identify all partitions relevant to a query, the master keeps a tree with the primary-key ranges of all partitions.
While re-partitioning, both nodes, the sending and receiving need to be accessed by queries to determine which node currently claims ownership over the data.
Therefore, when repartitioning starts, the master is updated first, keeping pointers to both, the old and new node.
After repartitioning, the old pointer is deleted.
\begin{figure*}[t!]%
	\centering
  \begin{subfigure}{0.9\columnwidth}
		\includegraphics[width=\textwidth, page=1]{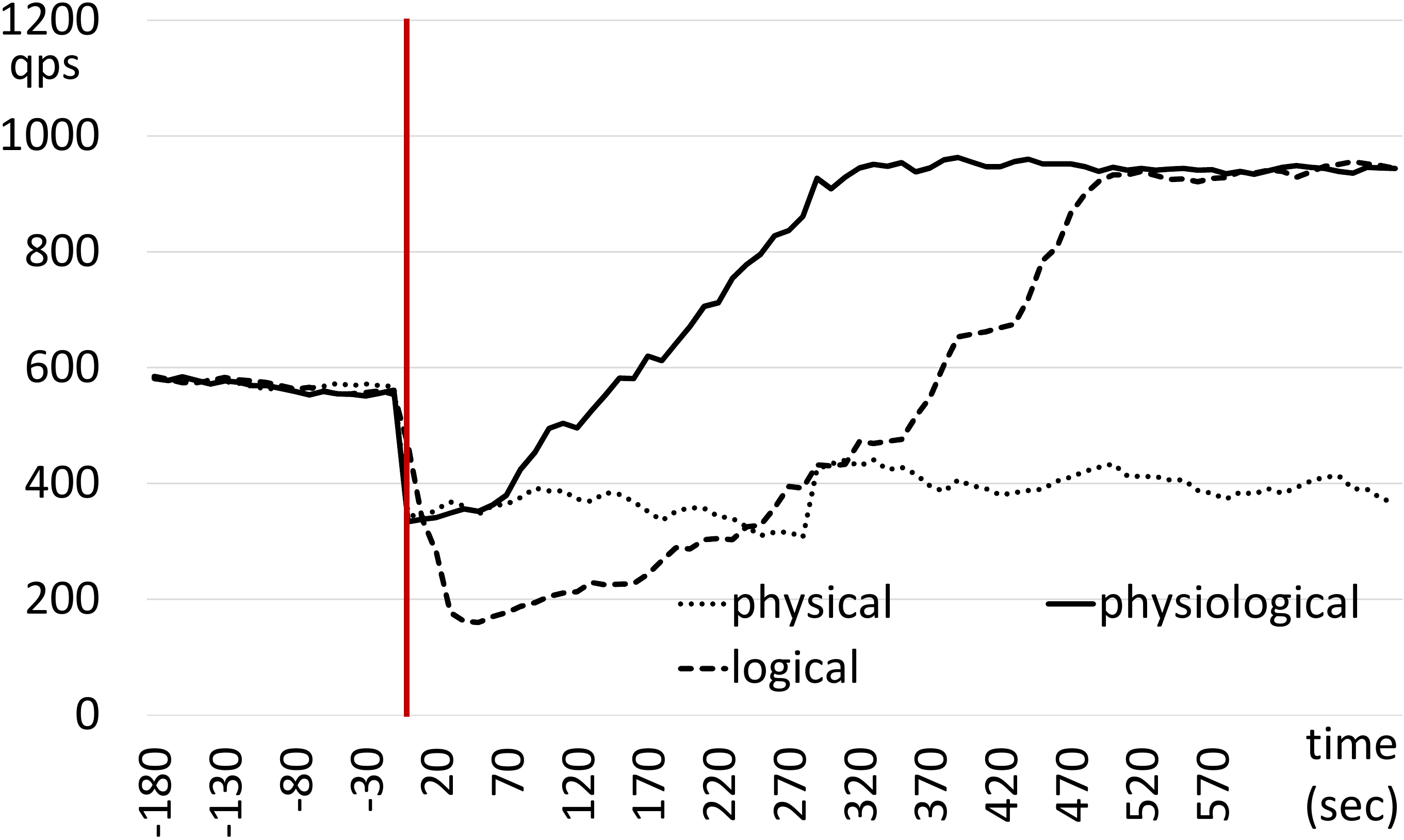}%
		\caption{Throughput  of the cluster}%
		\label{figure:partitioning:throughput}%
	\end{subfigure}%
  \begin{subfigure}{0.9\columnwidth}
		\includegraphics[width=\textwidth, page=2]{\imagedir/CIKM_Partitioning}%
		\caption{Avg. response time per query}%
		\label{figure:partitioning:responsetime}%
	\end{subfigure}%
	\\
  \begin{subfigure}{0.9\columnwidth}
		\includegraphics[width=\textwidth, page=3]{\imagedir/CIKM_Partitioning}%
		\caption{Power consumption of the cluster}%
		\label{figure:partitioning:powerconsumption}%
	\end{subfigure}%
  \begin{subfigure}{0.9\columnwidth}
		\includegraphics[width=\textwidth, page=4]{\imagedir/CIKM_Partitioning}%
		\caption{Energy consumption per query}%
		\label{figure:partitioning:energyconsumption}%
	\end{subfigure}%
	\caption{Benchmark results for various partitioning schemes under a TPC-C query mix}
	\label{figure:partitioning}
		\vspace{-0.2cm}
\end{figure*}

\textbf{Correctness:} Dynamic data migration must not alter the result of concurrent queries, hence, ACID properties must be maintained at all times.
Due to the copying/moving of records among partitions, transactions may access both copies or none of the copies by mistake.
In contrast to logical partitioning, rebalancing physiological partitions requires some modifications to MVCC to ensure correctness.
To show that transactions will behave correctly, we must provide proof of correctness for transactions at different starting times and distinguish between reading and writing accesses.

First, \textbf{transactions started prior to re-balancing} must be able to access old versions of the records.
Since the copies are kept until all old readers are finished, these transactions will always be able to read.
During rebalancing, a read lock is acquired on the old partition, ensuring that all writing transactions will finish until the partition is moved.
Newer transactions, arriving after locking the old partition, are either arriving at the new location and may write immediately, or are forced to wait for the copying to finish and are then redirected to the new partition.

Second, \textbf{transactions started after rebalancing} must not access old copies.
After updating the partition tree on the master node, new transactions will be redirected automatically to the new partition.
Before the update, transactions will behave identical to the first case and potential updates will be redirected to the new partition, where proper synchronization is enforced.

There is a small time window, where the partition tree on the master is not up-to-date and transactions may get routed to the old partition instead of the new one.
Therefore, the master keeps two pointers, indicating both, the new and old partition location and queries are advised to visit both, determining the correct location to use during execution.

\section{Experiments}
\label{section:Experiments}
To compare  energy consumption and performance impact of various partitioning schemes, we have evaluated all three implementations on our cluster with 
an OLTP workload.
In the following, we first describe the experimental setup, before we present our results.

\subsection{Experimental Setup}
For all experiments, we are using the dataset from the well-known TPC-C benchmark with a 
scale factor of 1,000. Hence, a thousand warehouses were generated on the cluster, consisting of about 100 GB of data.
Due to additional indexes and storage overhead, the final DB had approx. 200 GByte of raw data.

\textbf{Queries: }
We use queries from the TPC-C benchmark as workload drivers for our experiment. 
Because we do not compare our results with other TPC-C results,  we do not comply with the exact TPC-C benchmark specifications which are unessential to reveal differences of   partitioning schemes. 
For example, because our research prototype does not support multi-statement transactions with user interaction, we modified all queries to exclude (emulated) user interaction and to execute in "a single run" on the database.
Further, our benchmark deviates from other specifications, \eg, wait time and response time constraints, 60-day space requirements, and transactions mix definitions.

\textbf{Workload mix: }
In each experiment, we spawned a number of OLTP clients, sending queries to the DBMS.
Each client submits a randomly selected query at specified intervals.
If the query is answered, 
the next query is delayed until  
the subsequent interval similar to defined think times in the TPC-C specification.
Hence, the more OLTP clients and the lower the think time, the more utilization is generated.

By limiting the maximum throughput at the client side, this experiment differs from traditional benchmarking.
While established benchmarks as TPC-C use maximum throughput as the metric, we are interested in the DBMS fitness to adjust to a given workload by keeping throughput acceptable and optimize the number of nodes the DBMS is running on, and, thus, improve energy efficiency.

\textbf{Partitioning: }
As previously described, the limiting factor for dynamic repartitioning is migration cost, \ie, the performance impact and time 
to move data among nodes.
To estimate its impact on the cluster's elasticity, we have conducted a simple experiment: 

Starting with two nodes, hosting the data and processing queries, we instruct WattDB to perform a repartitioning of all tables and migrate 50\% of the records to two additional nodes.
We measure  response time, throughput, and power consumption of the cluster before, during and after the repartitioning.
We repeated the experiment on all three types of partitioning schemes, physical, logical, and physiological and compared the results.

\subsection{Results}
In this section, we present the results from our experiments.
Fig. \ref{figure:partitioning} illustrates  how---under the experiment sketched---query throughput, response time, power consumption, and energy use per query for the three benchmark runs evolve over time.
In each graph, the x-axis shows the time measured since initiating rebalancing in seconds.
At time $t\pm0$, the cluster was instructed to rebalance as previously described.
For $t < 0$, the results are more or less identical, because the initial configurations were identical for all experiments.
After starting repartitioning, measurements start to differ based on the selected partitioning scheme.
Because the same number of machines was used, power consumption is almost identical in all cases.

\textbf{Physical partitioning: }
Immediately after initiating rebalancing, query response times slightly increases and throughput reduces from about 600 to 400 qps, due to the network overhead of copying segments and the increased latency to access the now remote pages.
After moving 50\% of segments to new nodes (around $t+270$), query response time decreases, but does not recover to its old level.

With physical partitioning, segments are moved to another node, but are still "owned" by the same node.
Therefore, the partition can benefit from higher IOPS, due to the distribution, but suffers from increased network latency, because segment access now requires a remote call.

Referring to the measurements, we reason that physical partitioning---although easy to implement---is not usable for a dynamic cluster of DBMS nodes.
Applying this technique, we can distribute data among multiple disks, but the logical control of the data is stuck at the original node.
For this reason, storage segments have to be fetched from that node to access their records, which imposes additional latency.
Furthermore, without additional CPUs and main memory to help evaluating queries, scale-out can only be achieved at the storage layer.
Thus, physical partitioning is not useful for a fully dynamic DB cluster.

\textbf{Logical partitioning: }
Using logical partitioning, the control over a key range---together with the records---is transferred to another node.
Hence, moving records of a key range $[a - b)$ to another node requires the node to evaluate queries for that key range from now on.

The benchmark results on a logically partitioned cluster show an initial decline in query throughput (Fig. \ref{figure:partitioning:throughput}, at $t\pm0$).
Compared to the other schemes, logical partitioning exhibits the highest query response times when rebalancing (Fig. \ref{figure:partitioning:responsetime}).
After a significant amount of records has been relocated to other nodes, throughput and response times start to improve (at $t+170$) and quickly pass performance before repartitioning.
We explain the initial performance setback with the additional high system load due to table scan(s) and the network load for finding and moving records.
With parts of the data moved to another node, the original node does only have to manage the remainder of the data and the additional node takes part in query processing, doubling the number of CPUs and main memory available.

Hence, with logical partitioning, it is possible to add storage and processing power to the system, making it a better candidate for a dynamically adjusting cluster.
Yet, moving key ranges and scanning for data is time-consuming, compared to raw movement of physical segments.

\textbf{Physiological partitioning: }
The corresponding results for our benchmark on a physiologically partitioned cluster exhibit an initial decline in query performance (throughput and response times) similar to physical partitioning.
Similar to logical partitioning, performance in this approach recovery quickly and soon outperforms physical partitioning (around $t+250$), as soon as the majority of segments is transferred to the new nodes.
At this point, response times start to get lower than before, because all nodes can now participate in query processing.

In our experiments, physiological partitioning exhibited the lowest query runtimes and handles repartitioning events well, compared to the other approaches.
It provides fast adaption of data partitioning in a dynamic cluster and quickly compensates data migration overhead.
Overall, physiological partitioning delivers best energy efficiency and quickest adaptivity.
With this approach, we are combining the speed of data movement with the ability of transferring ownership of data.
The DBMS moves segments among nodes at the same speed as with physical partitioning.
As soon as segments arrive at the new node, they are incorporated in its index and the new node overtakes query processing.
Yet, immediately after the beginning of repartitioning, performance declines, further slowing query processing.
Therefore, physiological partitioning still shows drawbacks that need to be tackled.
\begin{figure}[h]
	\centering
	\includegraphics[width=0.8\columnwidth]{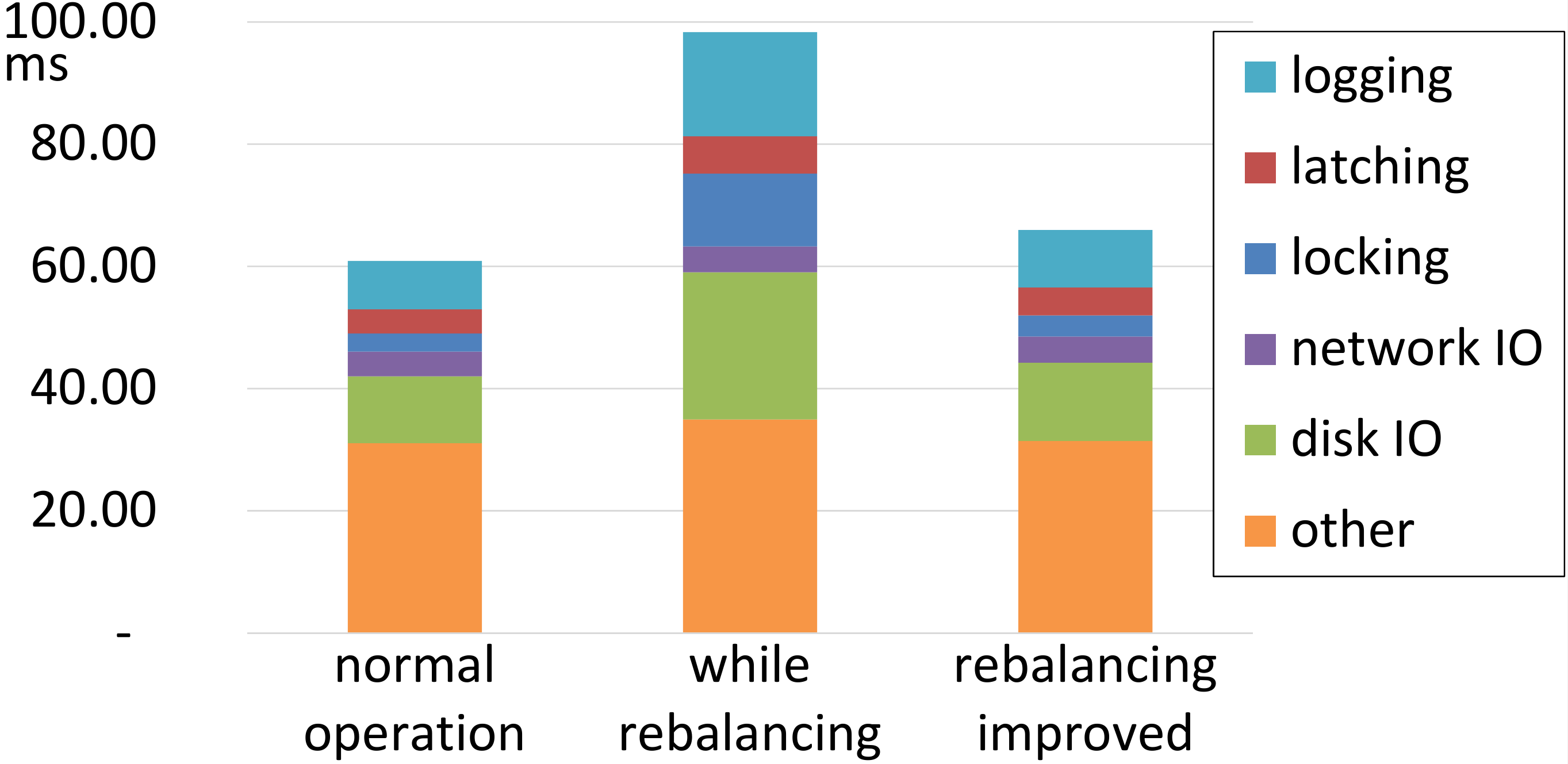}
	\vspace{-0.2cm}
	\caption{Impact factors on query runtime when rebalancing}
	\label{fig:physiological:overhead}
	\vspace{-0.4cm}
\end{figure}

\textbf{Physiological partitioning improved: }
From our first experiment on a physiological cluster, we experienced slow query response times during repartitioning.
We analyzed the performance setback and identified bottlenecks in the cluster.
In Fig. \ref{fig:physiological:overhead}, the major impact factors on query runtime on a physiologically partitioned cluster are shown.
On the left side, the graph shows a breakdown of time spent in various DBMS components when running queries.
On the right side, the same queries are running while the data is rebalanced to other nodes.
From the increase in runtimes, we can deduce that critical sections are disk I/O and locking.

Surprisingly, although repartitioning ships big chunks of data across the network, the time spent for network communication remains unchanged.
The findings indicate several bottlenecks:
First, locking partitions keeps queries waiting and thus, increases runtime.
Unfortunately, in the current implementation of the rebalancing operation, the lock is essential for data integrity.
Therefore, there is nothing we can do to mitigate locking overhead.

Second, rebalancing involves heavy I/O, competing with disk accesses of regular queries.
Reducing accesses to hard disk would therefore speed up query processing while repartitioning.
Additionally, we noticed more contention in the DB buffer due to a pile of waiting queries with latched pages and occupied pages needed for rebalancing (not shown in the figure).
More DRAM might reduce page thrashing and relieve the storage subsystem.

Lastly, as shown in Fig. \ref{fig:physiological:overhead}, logging takes significantly longer when rebalancing.
Since logging writes to disk as well, we conclude that the main bottleneck for repartitioning seems to be the bandwidth to the storage subsystem.

To mitigate excessive load on the cluster while balancing, we conducted a final experiment, where we powered up additional nodes to assist the present ones.
Since offloading OLTP query operators to remote nodes is not reasonable, we used the helper nodes for log shipping and provision of additional buffer space using rDMA\footnote{rDMA = remote direct memory access. A node's buffer size  is increased by including main memory from remote nodes}.
Accessing buffer pages from a remote memory includes network latency, but is still faster than flushing a page from the buffer and reading it back from disk when needed.
Especially "warm" data, that is not accessed frequently in the buffer (but frequent enough to justify keeping the page in memory), is a good candidate for rDMA buffering.

\begin{figure*}[t!]
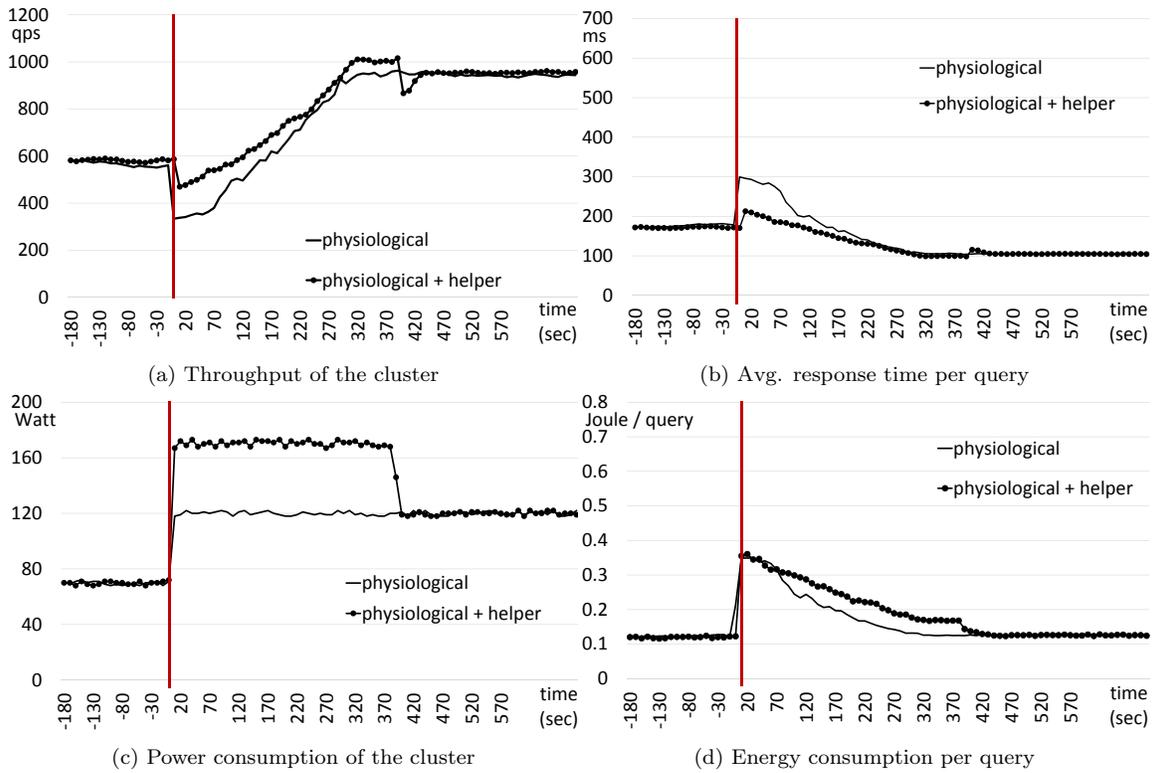
%
	\centering
  \begin{subfigure}{0.9\columnwidth}
		\includegraphics[width=\textwidth, page=5]{\imagedir/CIKM_Partitioning}%
		\caption{Throughput of the cluster}%
		\label{figure:partitioning2:throughput}%
	\end{subfigure}%
  \begin{subfigure}{0.9\columnwidth}
		\includegraphics[width=\textwidth, page=6]{\imagedir/CIKM_Partitioning}%
		\caption{Avg. response time per query}%
		\label{figure:partitioning2:responsetime}%
	\end{subfigure}%
	\\
  \begin{subfigure}{0.9\columnwidth}
		\includegraphics[width=\textwidth, page=7]{\imagedir/CIKM_Partitioning}%
		\caption{Power consumption of the cluster}%
		\label{figure:partitioning2:powerconsumption}%
	\end{subfigure}%
  \begin{subfigure}{0.9\columnwidth}
		\includegraphics[width=\textwidth, page=8]{\imagedir/CIKM_Partitioning}%
		\caption{Energy consumption per query}%
		\label{figure:partitioning2:energyconsumption}%
	\end{subfigure}%
	\caption{Improving the benchmark results for physiological partitioning}
	\label{figure:partitioning2}
	\vspace{-0.2cm}
\end{figure*}

The graphs in Fig. \ref{figure:partitioning2} plot the results in comparison with "standard" physiological partitioning.
At time $t\pm0$, when repartitioning started, two additional nodes were fired up to support the cluster.
After repartitioning was finished, the helper nodes were brought down again (around time $t+370$).
As the results confirm, including additional nodes increases power consumption (Fig. \ref{figure:partitioning2:energyconsumption}), but improves query response times (Fig. \ref{figure:partitioning2:responsetime}).
Overall, energy efficiency gets worse (more energy consumption per query, see Fig. \ref{figure:partitioning2:energyconsumption}), but, in turn, performance increases (Fig. \ref{figure:partitioning2:throughput}).

We conclude that adding nodes during rebalancing helps mitigate data shipment overhead at the cost of higher power consumption.
Therefore, after rebalancing, the additional nodes should be turned off again to improve energy efficiency of the cluster.

\section{Conclusion}
\label{section:Conclusion}
In this paper, we have evaluated different partitioning approaches for the use in an energy-proportional DBMS.
Our experiments identified drawbacks with physical and logical partitioning schemes and recommended physiological partitioning as best choice for dynamically repartitioning a database under load.

Yet, the experiments indicated that repartitioning an already stressed cluster imposes additional load and, thus, slows down query evaluation.
By activating additional nodes to support query processing, we were able to relieve some of the stress and to improve responsiveness of the system and, thus, we were able to trade energy efficiency for query performance.
\balance

\bibliographystyle{abbrv}
\bibliography{bibliography}

\begin{thebibliography}{10}

\bibitem{Oracle11gPartitioning}
{Oracle Database VLDB and Partitioning Guide 11g Release 1 (11.1) -
  Partitioning Concepts}, 2007.

\bibitem{DBLP:journals/computer/BarrosoH07}
L.~A. Barroso and U.~H{\"o}lzle.
\newblock {The Case for Energy-Proportional Computing}.
\newblock {\em IEEE Computer}, 40(12):33--37, 2007.

\bibitem{Bernstein:1983:MCC:319996.319998}
P.~A. Bernstein and N.~Goodman.
\newblock {Multiversion Concurrency Control---Theory and Algorithms}.
\newblock {\em ACM Trans. Database Syst.}, 8(4):465--483, Dec. 1983.

\bibitem{conf/cidr/BonczZN05}
P.~A. Boncz, M.~Zukowski, and N.~Nes.
\newblock {MonetDB/X100: Hyper-Pipelining Query Execution}.
\newblock In {\em CIDR}, pages 225--237, 2005.

\bibitem{Das:2011:SET:2521552}
S.~Das.
\newblock {Scalable and Elastic Transactional Data Stores for Cloud Computing
  Platforms}.
\newblock 2011.

\bibitem{Graefe:1994:VEP:627290.627558}
G.~Graefe.
\newblock {Volcano---An Extensible and Parallel Query Evaluation System}.
\newblock {\em IEEE Trans. on Knowl. and Data Eng.}, 6(1):120--135, Feb. 1994.

\bibitem{Graefe:2011}
G.~Graefe.
\newblock {Modern B-tree Techniques}.
\newblock {\em Foundations and Trends in Database}, 3(4):203--402, 2011.

\bibitem{KRAMER12}
C.~Kramer, V.~H\"ofner, and T.~H\"arder.
\newblock {Load Forcasting for Energy-Efficient Distributed DBMSs (in German)}.
\newblock {\em Proc. 42. GI-Jahrestagung 2012, LNI 208}, pages 397--411, 2012.

\bibitem{Lang:2012:TED:2350229.2350280}
W.~Lang and et~al.
\newblock {Towards Energy-Efficient Database Cluster Design}.
\newblock {\em PVLDB}, 5(11):1684--1695, 2012.

\bibitem{pandis2010data}
I.~Pandis, R.~Johnson, N.~Hardavellas, and A.~Ailamaki.
\newblock {Data-Oriented Transaction Execution}.
\newblock {\em Proc. VLDB Endowment}, 3(1-2):928--939, 2010.

\bibitem{Schall:2013B}
D.~Schall and T.~H\"{a}rder.
\newblock {Energy-Proportional Query Execution Using a Cluster of Wimpy Nodes}.
\newblock In {\em SIGMOD Workshops, DaMoN}, pages 1:1--1:6, 2013.

\bibitem{Schall:2013A}
D.~Schall and T.~H{\"a}rder.
\newblock {Towards an Energy Proportional Storage System using a Cluster of
  Wimpy Nodes}.
\newblock In {\em Proc. BTW, LNI 214}, pages 311--325, 2013.

\bibitem{SH-DASFAA2014}
D.~Schall and T.~H{\"a}rder.
\newblock {Approximating an Energy-Proportional DBMS by a Dynamic Cluster of
  Nodes}.
\newblock In {\em Proc. DASFAA}, pages 297--311, 2014.

\bibitem{Srinivasan:2000:OIT:645926.672004}
J.~Srinivasan and et~al.
\newblock Oracle8i index-organized table and its application to new domains.
\newblock In {\em Proc. VLDB}, pages 285--296, 2000.

\bibitem{DBLP:journals/vldb/TozunPJA13}
P.~T{\"o}z{\"u}n and et~al.
\newblock {Scalable and dynamically balanced shared-everything OLTP with
  physiological partitioning}.
\newblock {\em VLDB J.}, 22(2):151--175, 2013.

\bibitem{DBLP:conf/sigmod/TsirogiannisHS10}
D.~Tsirogiannis, S.~Harizopoulos, and M.~A. Shah.
\newblock {Analyzing the Energy Efficiency of a Database Server}.
\newblock In {\em Proc. SIGMOD}, pages 231--242, 2010.

\end{thebibliography}

\end{document}